\documentstyle[aps,prd,preprint]{revtex}

\begin{document}
\draft
\preprint{US-FT/1-96}

\title{Canonical quantization of the 
relativistic particle in static spacetimes}

\author{Alberto Saa}

\address{Departamento de F\'\i sica de Part\'\i culas,\\ 
Universidade de Santiago de Compostela,\\
E-15706, Santiago de Compostela, Spain }

\maketitle

\begin{abstract}
We perform the canonical quantization of a relativistic
spinless particle moving in a curved and static spacetime.
We show that the classical
theory already describes at the same time both particle and antiparticle. 
The analyses involves time-depending constraints and we are able
to construct the two-particle Hilbert space. 
The requirement of a static spacetime is necessary in order to have
a well defined Schr\"odinger equation and to avoid problems with
vacuum instabilities. The severe ordering ambiguities we found are in
essence the same ones of the well known non-relativistic case.
\end{abstract}

\vspace{1cm}
\begin{flushleft}
{PACS: 0350, 0365, 0420}
\end{flushleft}

\newpage

Many works have been devoted in the last 20 years to the study of the
quantization of relativistic particles (See, for instance, ref. \cite{GT}
and references therein). The canonical quantization of a relativistic
spinless particle in the presence of
background fields is a problem of especial interest.
For the case of electromagnetic background fields, canonical quantization
was performed for a large class of fields, mainly for that ones
 where no pair creation occurs, see \cite{GG} for further references. As to 
gravitational field case, in spite of the BRST quantization was done some
time ago\cite{JM}, the canonical quantization is still lacking. The great
difficulties are  the severe ordering ambiguities inherent to the
classical--quantum transition and the presence of time-depending
constraints. The quantization by Hamiltonian reduction
of a relativistic particle moving on some group manifolds was recently
considered in \cite{DOT}.

The purpose of this work is to present the canonical quantization of a
spinless relativistic particle moving in a static (pseudo)riemannian
manifold of dimension $D$. 
The canonical analysis involves time-depending constraints and as we will
see below, 
the classical theory already describes both particle and antiparticles at
the same time, in spite of the original lagrangian is a single particle one.
We construct the two-particle Hilbert space $\cal H$, which mimics some
properties of its Minkowskian counterpart\cite{GT1}.
The severe ordering ambiguities that we found are in
essence the same ones of the well known non-relativistic case, and we will
consider the most general hermitean and invariant Hamiltonian operator.
We start the analysis with the
lagrangian formulation of a spinless relativistic particle moving in
a static manifold, and in order to perform the canonical
quantization we go to hamiltonian formulation by following standard
steps of the theory of constrained systems\cite{GT,HT}.

First, we shall explain precisely what means a static manifold. We call
a riemannian manifold static 
if\cite{F}: (a) There is a timelike Killing
vector field, and (b) There is a family of spacelike surfaces orthogonal to the
Killing vector everywhere. These requirements are equivalent to, in an
appropriate coordinate system where $x^0$ is timelike, the following 
restrictions on the metric $g_{\mu\nu}$ of the manifold
\begin{eqnarray}
\label{restr}
&\rm{(a)\ }&g_{\mu\nu}(x){\rm\ is\ independent\ of\ }x^0, \nonumber  \\
&\rm{(b)\ }&g_{0j}(x) = 0.    
\end{eqnarray}
It is assumed hereafter 
that Greek indices run over ($0,D-1$), and Roman ones over
($1,D-1$). We adopt the conventions of ref. \cite{GT}, and in particular the
metric has signature $(1,-1,...,-1)$. We assume also that 
$g=|\det\{g_{\mu\nu}\}|$. It may seem that the conditions
(\ref{restr}) are too restrictive, and we remind that physically
relevant examples as the exterior regions of Schwarzschild and 
Reissner-Nordstr\"om solutions and
the Rindler metric obey (\ref{restr}). These restrictions are used also
in many of the path-integral approaches to the problem, mainly when
initial value problems are treated\cite{Fe}.
When the space-time  obeys (\ref{restr}), the quantity 
$\sqrt{g_{00}(x)}=\left(\sqrt{g^{00}(x)}\right)^{-1}$ 
is called lapse function, 
because it measures the distance
between the spacelike surfaces $x^0$ and $x^0 + dx^0$ constants.
We will discuss latter the necessity
and the actual role of such restrictions.

We  start with the action of a relativistic spinless particle in a static
manifold,
\begin{eqnarray}
\label{act}
S &=& -m\int ds = -m\int L d\tau, \nonumber \\
L &=& -m\sqrt{g_{\alpha\beta}(x) \dot{x}^{\alpha}\dot{x}^{\beta}}, \ \ \ \ \ 
\dot{x}^{\alpha} = \frac{dx^\alpha}{d\tau}.
\end{eqnarray}
The action (\ref{act}) is invariant under the reparameterizations 
$\tau \rightarrow f(\tau)$ with $\dot{f}>0$. Due to this the lagrangian
$L$ is singular, and we get primary constraints when  going to the 
hamiltonian formalism. To see it, let us introduce the canonical momenta
\begin{equation}
\label{can}
\pi_\mu = \frac{\partial L}{\partial \dot{x}^\mu} = 
- \frac{m\dot{x}_\mu}{\sqrt{\dot{x}^2}},
\end{equation}
and one can easily check that the constraint 
$g^{\mu\nu}\pi_{\mu}\pi_{\nu} = m^2$ holds, 
which we write for convenience in the equivalent form
\begin{equation}
\label{constr1}
\Phi^{(1)} = {\sqrt{g_{00}}}{\sqrt{m^2 - g^{ij}\pi_i\pi_j }}  - |\pi_0| =0 .
\end{equation}
The situation here is similar to the flat space case\cite{GT}. From
(\ref{can}) we can express only the velocities $\dot{x}^i$ and the sign
of $\dot{x}^0$ by means of the variables $\pi_i$, $\lambda=|\dot{x}^0|$, and
$\xi = - {\rm sign\ } \pi_0$,
\begin{equation}
\dot{x}_i = -\frac{\pi_i\sqrt{g_{00}}\lambda}{\sqrt{m^2-g^{ij}\pi_i\pi_j }},
\ \ \ \ {\rm sign\ } \dot{x}^0 = \xi.
\end{equation}
The modulus of $\dot{x}^0$, $\lambda$, can not be expressed by means of
(\ref{can}).
The hamiltonian $H^{(1)}$ can now be construct calculating
$\pi_\mu\dot{x}^\mu -L$, and we obtain
\begin{equation}
H^{(1)} = \lambda \Phi^{(1)}. 
\end{equation}

One can follow Dirac procedure and verify that no more constraints arise and
that $\lambda$ remains undetermined. The model involves only one first class
constraint, and to continue the analysis one needs to choose a 
 gauge fixing condition\cite{GT,HT}. We use
\begin{equation}
\label{G}
\Phi^{G} = x^0 - \xi\tau.
\end{equation}
From the condition of conservation of this gauge choice in $\tau$ one 
can determine $\lambda$. To avoid $\tau$-depending constraints,
we make the canonical transformation that leads ${x}^{0'} = x^0 - \xi\tau$
and leaves all the other canonical variables unchanged. In the new variables
 the gauge fixing is given by $\Phi^G = x^{0'} = 0$. One can check that
such canonical transformation is defined by the generating function
$W = \pi'_\mu {x}^{\mu} + \tau |\pi'_0|$, and that the hamiltonian
transforms as
\begin{equation}
\label{h1}
H^{(1)'} = H^{(1)} + \frac{\partial W}{\partial\tau} = 
\sqrt{g_{00}}\sqrt{m^2-g^{ij} \pi_i \pi_j } + (\lambda - 1)\Phi^{(1)}.
\end{equation}
The constraints $\Phi = (\Phi^G, \Phi^{(1)})$ form a set of second class ones
and are of special form\cite{GT},
and we can use them to eliminate the variables $x^0$ and $|\pi_0|$. We can
check also that the Dirac brackets of the physical variables $x^i,\pi_i,$
and $\xi$ with respect to the constraints $\Phi$ reduce to the ordinary
Poisson ones.
The restriction of (\ref{h1}) on the constraint surface gives the physical
hamiltonian $H$, which describe the dynamics of the physical variables,
\begin{eqnarray}
\label{equ}
H &=& \sqrt{g_{00}}{\sqrt{ m^2-g^{ij}\pi_i\pi_j}},\nonumber \\
\dot{x}^i &=& \{x^i,H \}, \\
\dot{\pi}_i &=& \{\pi_i,H \}, \nonumber\\
\dot{\xi} &=&0 \nonumber .
\end{eqnarray}
As in the flat space case, the variable $\xi$, that assumes the values
$\pm 1$, is a constant of motion. We can interpret the variable $\xi$
by introducing an external electromagnetic field. The situation is analogous
to the flat space case\cite{GT}, and by introducing a magnetic field
and comparing the trajectories one  concludes
that $\xi = 1$ and $\xi =-1$ correspond respectively to the trajectories of
particles and antiparticles. The canonical quantization will confirm such
conclusion. 

Now we can proceed with the quantization of the system described by
(\ref{equ}). The only non vanishing commutator for the Schr\"odinger operators
$\hat{x}^i,\hat{\pi}_i$, and $\hat{\xi}$ is
\begin{equation}
\left[\hat{x}^j,\hat{\pi}_k\right] = i\delta^j_k.
\end{equation}
By analogy with the classical theory, let us assume that the operator
$\hat{\xi}$ has eigenvalues ${\xi}=\pm 1$. We introduce the Hilbert
space $\cal H$, whose elements $\Psi$ are complex two-components columns
\begin{equation}
\Psi(x^\mu) = \left( 
\begin{array}{c}
\psi_+(x^\mu) \\
\psi_-(x^\mu)
\end{array} 
\right),
\end{equation}
and the invariant inner product is given by
\begin{equation}
\label{inner}
\left< \Psi^1,\Psi^2\right> = \int d^D x \sqrt{g} \Psi^{1^\dagger}\Psi^2.  
\end{equation}
We choose for our Schr\"odinger operators the following representation
\begin{eqnarray}
\label{repr}
\hat{\xi} &=& \left( 
\begin{array}{cc}
1 & 0 \\
0 & -1
\end{array}\right), \nonumber \\
\hat{x}^k &=& x^k{\bf I} , \\
\hat{\pi}_k &=& -ig^{-\frac{1}{4}} 
\partial_k g^{\frac{1}{4}} {\bf I}, \nonumber 
\end{eqnarray}
where {\bf I} is the unit $2\times 2$ matrix. 
All these operators are hermitean with
respect to the inner product (\ref{inner}) (We assume for simplicity that
$\psi_+$ and $\psi_-$ have compact support).

The dynamics of the physical states $\Psi_{\rm ph} \in \cal H$ 
are described by the Schr\"odinger equation
\begin{equation}
\label{schr}
i\frac{\partial}{\partial\tau}\Psi_{\rm ph} = \hat{H}\Psi_{\rm ph},
\end{equation}
where $\hat{H}$ is the quantum counterpart of the hamiltonian $H$ in
(\ref{equ}). It is clear that the determination of $\hat{H}$ is plagued
with ordering ambiguities. It is more convenient to our purposes to
 introduce the physical time $x^0$ in (\ref{schr}). We can do it by exploring
(\ref{G}), and we have
\begin{equation}
\label{schr1}
i\frac{\partial}{\partial x^0}\Psi_{\rm ph} = \hat{\xi}\hat{H}\Psi_{\rm ph}.
\end{equation}
These last equations deserve some comments. It is here that for the
first time the necessity of the restriction to a static spacetime arises.
It is the existence of a timelike Killing vector that makes possible
 to write (\ref{schr}) as (\ref{schr1}). Also,
in order to have a well-defined Cauchy problem to the equation
(\ref{schr1}) we need to have spacelike surfaces orthogonal
to the timelike Killing vector. In fact we need also that this surfaces be
Cauchy surfaces, or what is equivalent that the space-time be 
globally hiperbolic\cite{F}.

For our purposes, it is more convenient to consider now the second order
equation obtained by applying $ig^{00}\frac{\partial}{\partial x^0}$
to (\ref{schr1}),
\begin{equation}
\label{KG}
K\Psi_{\rm ph} = 
\left(g^{00}\partial_0^2 + \tilde{H}^2\right)\Psi_{\rm ph} = 0,
\end{equation}
where $\tilde{H}^2 = g^{00}\hat{H}$. Each component of the state
vector $\Psi_{\rm ph}$ will obey (\ref{KG}). Now, it turns out that it is
more easy to write down a general form for $\tilde{H}^2$ from 
(\ref{KG}), than to do for $\hat{H}$ from (\ref{schr1}).
There are examples in
the literature where some ordering ambiguities can be
 solved by iterating the
relevant operators\cite{GS}.
It is known\cite{DW} that to define a
time-invariant scalar product for the solutions of (\ref{KG}) the operator
$K$ must be hermitean with respect to (\ref{inner}). Also,
since the state vectors $\Psi_{\rm ph}$ are assumed to be scalars
 under coordinate
transformations, the equation (\ref{KG}) must be covariant.
These two conditions, the classical expression for $H$, and the
requirement that only terms up to $\hbar^2$ order should be present in
the Hamiltonian $\tilde{H}^2$ lead to the following general expression:  
$\tilde{H}^2 = \hat{\pi}^2 + m^2 + \lambda R$, where
$\hat{\pi}^2=g^{-\frac{1}{4}}\hat{\pi}_i
\sqrt{g}g^{ij}
\hat{\pi}_j g^{-\frac{1}{4}}$, $R$ is the scalar of curvature, and 
$\lambda$ is a real number.
The equation (\ref{KG}) becomes
\begin{equation}
\label{KG1}
K\Psi_{\rm ph} = \left( \frac{1}{\sqrt{g}}
\partial_\mu\sqrt{g}g^{\mu\nu}\partial_\nu
 +m^2 + \lambda R \right)\Psi_{\rm ph} = 0,
\end{equation}
which is the standard generalization of the Klein-Gordon equation in a 
curved manifold. There is a vast literature about the choice of the
constant $\lambda$, but we do not pay attention to it because 
its value does not affect our analysis and
results. The
time independent scalar product for the $\Psi_{\rm ph}$ in this case
is given by
\begin{equation}
\label{i1}
\left< \Psi_{\rm ph}^1, \Psi_{\rm ph}^2\right>_{\rm ph} = 
\int d^{D-1}x \sqrt{\tilde{g}} \sqrt{g^{00}}\left( 
{\Psi_{\rm ph}^{1^\dagger}} \partial_0 \Psi_{\rm ph}^2 - 
\left(\partial_0 {\Psi_{\rm ph}^{1^\dagger}}\right) \Psi_{\rm ph}^2
\right),
\end{equation}
where $d^{D-1}x\sqrt{\tilde{g}}$ is the invariant 
volume element in one of the
spacelike surfaces $x^0={\rm const,}$ 
and as in the flat space case (\ref{i1}) is not positive defined. 

Since the role of the operator $\hat{\xi}$ 
is realized only in the Schr\"ondiger
equation ($\hat{\xi}^2 = \bf I$), it would be of great interesting to continue
with the analysis of (\ref{schr1}). To this purpose we will restrict
ourselves to  ultrastatic spacetimes\cite{F},
what means that besides of the restrictions (\ref{restr}), we have also
$\sqrt{g_{00}}=1$. 
This restriction is convenient to avoid other ordering ambiguities with
the lapse function and it does not affect the role played by the 
operators $\hat{\xi}$.
With such a metric the distance between the
two surfaces labeled by $x^0=\rm $ and $x^0 + dx^0$
is independent of $x^i$. In this case we define formally 
$\hat{H} = \sqrt{\hat{\pi}^2 + m^2 + \lambda R}$, 
to proceed with the analysis
of (\ref{schr1}). The variable $x^0$ is perfectly separated from the
$x^i$ in (\ref{schr1}), what allows us to write its solutions 
$\Psi_{\rm ph}$ as
\begin{equation}
\Psi_{\rm ph}(x^\mu) = \left( 
\begin{array}{c}
e^{-i\omega x^0} f(x^i) \\
e^{i\omega x^0} f(x^i)
\end{array}
\right), \ \ \ \ 
\sqrt{\hat{\pi}^2 + m^2 + \lambda R}f(x^i) = \omega f(x^i).
\end{equation}
Now it is clear that the eigenstates of $\hat{\xi}$,
$\Psi_{\rm ph}^+ = e^{-i\omega x^0} f(x^i)\left(
\begin{array}{c}
1 \\
0
\end{array}\right)$ and  
$\Psi_{\rm ph}^- = e^{i\omega x^0} f(x^i)\left(
\begin{array}{c}
0 \\
1
\end{array} \right)$, are eigenstates of the operator $i\partial_0$ with
eigenvalues $\omega$ and $-\omega$ respectively. Written in a coordinate free
way we have
\begin{equation}
\label{lie}
i{\cal L}_{v} \Psi_{\rm ph}^\pm = \pm \omega \Psi_{\rm ph}^\pm,
\end{equation}
where ${\cal L}_{v}$ stands for the Lie derivative along the
timelike Killing vector $v$. One recognize (\ref{lie})
as the covariant separation in parts of positive
and negative frequencies of the wave function $\Psi_{\rm ph}$\cite{BD},
what confirms in the quantum dynamics the classical interpretation that
$\xi$ distinguishes between particles and antiparticles. We know also
that (\ref{lie}) guarantees that the vacuum is stable\cite{BD}, avoiding pair
creation and annihilation, 
in agreement with the fact that $\xi$ is a constant of motion.
In a general spacetime without a timelike Killing vector, the separation
(\ref{lie}) is not possible, and we will have inequivalent vacua connected by
non-trivial 
Boguliubov transformations. In such case, we could not use the gauge
choice (\ref{G}), and analysis would become extremely more complicated.

It is interesting to note also that in an ultrastatic spacetime one can
define a positive defined time-invariant inner product for the
$\Psi_{\rm ph}$ as
\begin{equation}
\label{inn1}
\left< \Psi_{\rm ph}^1, \Psi_{\rm ph}^2\right>_{\rm ph} = 
\int d^{D-1}x \sqrt{\tilde{g}}  
\Psi_{\rm ph}^{1^\dagger}  \Psi_{\rm ph}^2 ,
\end{equation}
and we can check that all the physical operators are hermitean with
respect to (\ref{inn1}).

As the conclusion, we stress that in a static spacetime is possible
to perform the canonical quantization of the spinless relativistic
particle getting the Hilbert space $\cal H$ describing a two-particle
quantum mechanics. Both particle and antiparticle are already present
at the classical level, corresponding to trajectories 
labeled by the two possible values of $\xi$.

 The author wishes
to thank CNPq for the financial support.

\end{document}